\documentclass[12pt]{iopart}

\usepackage{graphicx}

\usepackage{psfrag}

\eqnobysec

\begin{document} 
\title[Dispersion relations and speeds of 
sound]{Dispersion relations and speeds of sound in special sectors for the 
integrable chain with alternating spins} 
\author{B - D D\"orfel\footnote{E-mail:
doerfel@qft2.physik.hu-berlin.de} and St Mei\ss ner\footnote{E-mail:
meissner@qft2.physik.hu-berlin.de}} \address{Institut f\"ur Physik,
Humboldt-Universit\"at , Theorie der Elementarteilchen\\
Invalidenstra\ss e 110, 10115 Berlin, Germany}

\begin{abstract} 
Based on our previous analysis \cite{doerfel3} of the anisotropic integrable
chain consisting of spins $s=\frac{1}{2}$ and $s=1$ we compare the dispersion 
relations for the sectors with infinite Fermi zones. Further we calculate the
speeds of sound for regions close to sector borders, where the Fermi radii 
either vanish or diverge, and compare the results.
\end{abstract} 

\pacs{75.10 JM, 75.40 Fa}

\maketitle 
\section{Introduction} 
The integrable spin chain $XXZ(\frac{1}{2},1)$ constructed in 1992 by de Vega 
and Woynarovich \cite{devega} shows a rich physical structure with different 
ground states depending on the anisotropy parameter and the two coupling 
constants. In our previous papers \cite{doerfel3}, \cite{meissner}, 
\cite{doerfel1} and \cite{doerfel2} we have analyzed this structure and 
calculated several important quantities. In this paper we deal mainly with 
dispersion relations and speeds of sound.

The paper is organized as follows. After having reviewed the definitions in 
section 2 we found it necessary to remind the reader results of our last paper 
\cite{doerfel3} concerning the phase diagram. In the same section the 
dispersion relations for all sectors with infinite Fermi zones only are 
compared. In section 4 we present the calculations for the speeds of sound 
obtained either by power expansion or Wiener-Hopf technique for small or large
Fermi radii respectively.

Our conclusions are contained in section 5. Some useful definitions are 
compiled in an appendix.

\section{Description of the model}
We refer the reader to papers \cite{devega} and \cite{meissner} for the basics 
of the model.

Our Hamiltonian of a spin chain of length $2N$ is given by
\begin{equation}\label{ham}
{\cal H}(\gamma) = \bar{c} \bar{\cal H}(\gamma) + \tilde{c} \tilde{\cal H}
(\gamma),
\end{equation}
with the two couplings $\tilde{c}$ and $\bar{c}$. The anisotropy parameter 
$\gamma$ is limited to $0<\gamma<\pi/2$. For convenience we repeat the Bethe 
ansatz equations (BAE), the magnon energies and momenta and the spin projection.
\begin{equation}\label{bae}
\fl \left( \frac{\sinh(\lambda_j+i\frac{\gamma}{2})}{\sinh(\lambda_j-i
\frac{\gamma}{2})}
\frac{\sinh(\lambda_j+i\gamma)}{\sinh(\lambda_j-i\gamma)} \right)^N =
-\prod_{k=1}^{M}\frac{\sinh(\lambda_j-\lambda_k+i\gamma)}{\sinh(\lambda_j-
\lambda_k-i\gamma)},\qquad j=1\dots M,
\end{equation}
\begin{eqnarray}\label{en}
E = \bar{c} \bar{E} + \tilde{c} \tilde{E},
\nonumber\\
\bar{E} = - \sum_{j=1}^{M} \frac{2\sin\gamma}
{\cosh2\lambda_j - \cos\gamma},
\nonumber\\
\tilde{E} = - \sum_{j=1}^{M} 
\frac{2\sin2\gamma}{\cosh2\lambda_j - \cos2\gamma},
\end{eqnarray}
\begin{equation}\label{mom}
P =\frac{i}{2}\sum_{j=1}^{M} \left\{ \log \left(\frac{\sinh(\lambda_j+i\frac{
\gamma}{2})}{\sinh(\lambda_j-i\frac{\gamma}{2})} \right) + 
\log \left( \frac{\sinh(\lambda_j+i\gamma)}{\sinh(\lambda_j-i\gamma)} \right) 
\right\},
\end{equation}
\begin{equation}\label{spin}
S_z = \frac{3N}{2} - M.
\end{equation}

\section{Phase structure and dispersion relations}
In section 4 of \cite{doerfel3} we have carried out the complete analysis of 
the ground state structure based on the solution of the thermodynamic Bethe 
ansatz. We will now review the main results.

Using different sets of special anisotropy points we were able to prove that 
(for $0<\gamma<\pi/2$) only three kinds of strings occur in the ground state. 
The picture changes remarkably at $\gamma=\pi/3$ and $\gamma=2\pi/5$. All 
possible sectors are compiled in table 1 taken from \cite{doerfel3}. To keep 
our notation consistent we found it useful to call the sector already discussed
in \cite{devega} sector 0.
\begin{table}
\begin{center}
\begin{tabular}{|c||c|}\hline
0 & $(1,+)^{\infty}$, $(2,+)^{\infty}$\\ \hline
I & $(1,-)^{\infty}$, $(1,+)^{f,0}$\\ \hline
II & $(1,-)^{\infty}$, $(1,+)^{\infty}$\\ \hline
III & $(1,-)^{\infty}$, $(1,+)^{f,\infty}$\\ \hline
IV  & $(1,-)^{\infty}$\\ \hline
V & $(1,-)^{\infty}$, $(2,+)^{f,0}$\\ \hline
VI & $(1,-)^{f,\infty}$, $(2,+)^{\infty}$\\ \hline
\end{tabular}
\end{center}
\caption{\label{t}All sectors appearing in the phase diagram. 
Upper indices indicate infinite and finite Fermi zones. In the latter case, 
the second index distinguishes, wether the filling starts at $\lambda=0$ or 
$\lambda=\infty$.}
\end{table}
In the $(\tilde{c},\bar{c})$-plane the sector borders are straight lines 
starting from the origin. Moving counterclockwise from the $\tilde{c}$-axis the
sectors follow in sequence 0, I, IV, V if $0<\gamma\leq\pi/3$. For 
$\pi/3<\gamma<2\pi/5$ the sequence is 0, II, I, IV, VI and for 
$2\pi/5<\gamma<\pi/2$ one has 0, II, III, IV, VI. In case $\gamma=2\pi/5$ 
sector II is followed by IV and neither I or III occur. For details see figure 
1 in \cite{doerfel3}.

To compare the dispersion relations in the sectors 0, II and IV it is 
convenient to work with the functions
\begin{eqnarray}\label{p}
p(\lambda,\alpha,\beta)=\arctan \frac{\sinh(\pi\lambda/\beta)}
{\cos(\pi\alpha/2\beta)}
\end{eqnarray}
and
\begin{eqnarray}\label{g}
g(\lambda,\alpha,\beta)=\frac{4\pi}{\beta} 
\frac{\cos(\pi\alpha/2\beta)\cosh(\pi\lambda/\beta)}
{\cosh(2\pi\lambda/\beta)+\cos(\pi\alpha/2\beta)}.
\end{eqnarray}
While $p$ is odd, $g$ is an even function of $\lambda$. Both are connected by
\begin{eqnarray}\label{conn}
\frac{d}{d\lambda} p(\lambda,\alpha,\beta)=\frac{1}{2} g(\lambda,\alpha,\beta).
\end{eqnarray}
By straightforward calculation one may prove the relation
\begin{eqnarray}\label{rel}
\fl
\sin\left(\frac{\pi}{2}+p(\lambda,\alpha,\beta)\right) 
\sqrt{\cos^2 p(\lambda,\alpha,\beta) + \frac{\sin^2 p(\lambda,\alpha,\beta)}
{\cos^2(\pi\alpha/2\beta)}}=\frac{\beta}{2\pi} g(\lambda,\alpha,\beta).
\end{eqnarray}
We remark that in \cite{doerfel1} we used a function $g(\lambda,\alpha)$ 
related by
\begin{eqnarray}
g(\lambda,\alpha,\pi-\gamma)=g(\lambda,\alpha).
\end{eqnarray}

Let us first consider sector 0. The results are given in \cite{devega} taking
into account our definition of momentum in equation (\ref{mom}). One has two 
elementary physical excitations, the holes in the distributions of the $(1,+)$-
and $(2,+)$-strings respectively. We label them by index $i$, $i=1,2$.
\begin{eqnarray}\label{holemom1}
p_h^i(\lambda=\lambda_h^i)=\frac{\pi}{4}+\frac{1}{2} p(\lambda,0,\gamma),
\end{eqnarray}
\begin{eqnarray}\label{holeen1}
\varepsilon_h^i(\lambda)=c^i\frac{1}{2} g(\lambda,0,\gamma)
\end{eqnarray}
with $c^1=\bar{c}$ and $c^2=\tilde{c}$.

From equation (\ref{rel}) the dispersion relations follow
\begin{eqnarray}\label{disp1}
\varepsilon_h^i(p_h^i)=\frac{2\pi}{\gamma} c^i \frac{\sin 2p_h^i}{2}.
\end{eqnarray}
One may say that the dispersion relations diagonalize in the coupling 
constants. This fact changes in the other two sectors.

Next we look at sector IV which contains the case of negative couplings. The 
exact border lines are given in \cite{doerfel2}. The elementary physical 
excitations are holes in the distribution of the $(1,-)$-strings. We quote the 
result from our paper \cite{meissner}, equations (4.5) and (4.6)
\begin{eqnarray}\label{holemom2}
p_h(\lambda)=\frac{3\pi}{2}+\frac{1}{2} p(\lambda,0,\pi-\gamma) +
p(\lambda,\gamma,\pi-\gamma),
\end{eqnarray}
\begin{eqnarray}\label{holeen2}
\varepsilon_h(\lambda)=-\frac{\bar{c}}{2} g(\lambda,0,\pi-\gamma) -
\tilde{c} g(\lambda,\gamma,\pi-\gamma).
\end{eqnarray}
The dispersion relation cannot be obtained in a closed form. One has to invert
the monotone function (\ref{holemom2}) and to substitute it in equation 
(\ref{holeen2}). Numerical calculations show an sine-like behaviour, see the 
broken lines in figure 2 of \cite{meissner}.

\begin{figure}
\psfrag{gamma=3_7}{$\gamma=3\pi/7$}
\psfrag{gamma=3_8}{$\gamma=3\pi/8$}
\psfrag{gamma=2_5}{$\gamma=2\pi/5$}
\includegraphics [width=\columnwidth]{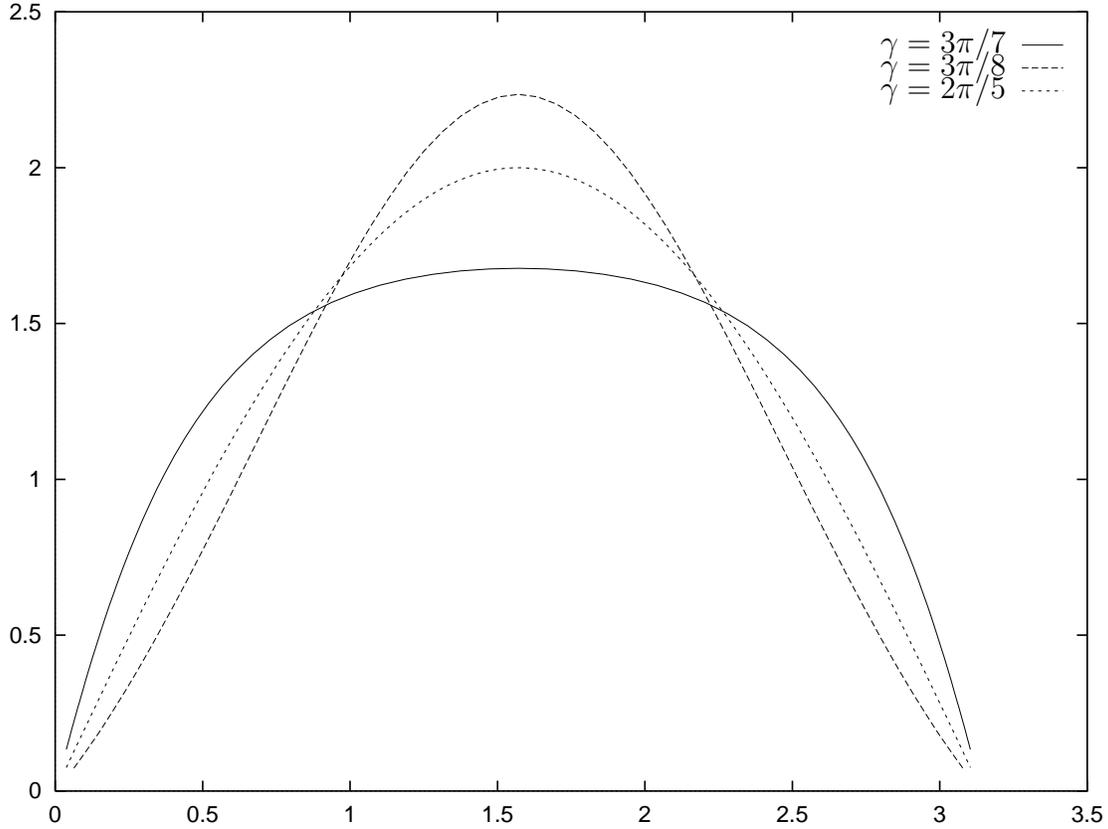}
\caption{\label{figdisp4}Dispersion relations for the holes in the 
$(1,+)$-string distributions in sector II for $\bar{c}=-5\tilde{c}=1$ and 
different $\gamma$.}
\end{figure}

In \cite{doerfel3} we found the new sector II, existing only for $\gamma>\pi/3$
(for the exact border lines see next section). Here the excitations are 
$(1,+)$- and $(1,-)$-string holes. The dispersion law for the $(1,-)$-string 
holes is simple:
\begin{eqnarray}\label{holemom3}
p_h^{1-}(\lambda)=\frac{\pi}{4}+\frac{1}{2} p(\lambda,0,\pi-2\gamma),
\end{eqnarray}
\begin{eqnarray}\label{holeen3}
\varepsilon_h^{1-}(\lambda)=-\frac{\tilde{c}}{2} g(\lambda,0,\pi-2\gamma),
\end{eqnarray}
\begin{eqnarray}\label{disp3}
\varepsilon_h^{1-}(p_h^{1-})=-\frac{2\pi\tilde{c}}{\pi-2\gamma}
\frac{\sin 2 p_h^{1-}}{2}.
\end{eqnarray}
To write down the case of $(1,+)$-strings it is necessary to introduce a new
function $d_1(\lambda)$
\begin{eqnarray}\label{d1}
d_1(\lambda)=\frac{1}{2\pi}\int_{-\infty}^{\infty}e^{i\omega\lambda}
\frac{\cosh(\omega(\pi-3\gamma)/2)d\omega}{2\cosh(\omega(\pi-2\gamma)/2)
\cosh(\omega\gamma/2)}
\end{eqnarray}
already used in \cite{doerfel3}. We further define another function 
$e_1(\lambda)$ by
\begin{eqnarray}\label{e1}
\frac{d}{d\lambda} e_1(\lambda)= d_1(\lambda)\quad\mbox{and}\quad e_1(0)=0.
\end{eqnarray}
\begin{eqnarray}\label{holemom4}
p_h^{1+} = \frac{\pi}{2} + \frac{1}{2} p(\lambda,0,\gamma) + 2\pi e_1(\lambda),
\end{eqnarray}
\begin{eqnarray}\label{holeen4}
\varepsilon_h^{1+} = \bar{c}\frac{1}{2} g(\lambda,0,\gamma) +  2\pi\tilde{c}d_1(\lambda).
\end{eqnarray}
We have calculated the dispersion law numerically and show the results in 
figure \ref{figdisp4}.
It is interesting to look closer at the point $\gamma=2\pi/5$, where the 
function $d_1(\lambda)$ simplifies and becomes $1/4\pi g(\lambda,0,\gamma)$,
resulting in the simple dispersion law
\begin{eqnarray}\label{disp4simple}
\varepsilon_h^{1+} = \frac{2\pi}{\gamma} \frac{\bar{c}+\tilde{c}}{2}
\sin p_h^{1+}.
\end{eqnarray}
All dispersion laws possess a symmetry connected with the properties of the 
functions $p_h(\lambda)$ and $\varepsilon_h(\lambda)$:
\begin{eqnarray}\label{symmom}
p_h(\lambda)=p_0 + \bar{p}(\lambda)\quad\mbox{and}\quad 
\bar{p}(-\lambda)=-p(\lambda),
\end{eqnarray}
\begin{eqnarray}\label{symen}
\varepsilon_h(\lambda)=\varepsilon_h(-\lambda).
\end{eqnarray}
One then derives easily
\begin{eqnarray}\label{symdisp}
\varepsilon_h(p_h)=\varepsilon_h(2 p_0 - p_h).
\end{eqnarray}
In addition one has $0\leq p_h\leq 2p_0$. The constants $2 p_0$ are different,
so in sector 0 they are $\pi/2$ for both excitations, the same is true for the 
negative parity excitation in II. The other (positiv parity) excitation in II
has $2 p_0=\pi$, while in IV one has $2 p_0=3\pi/2$.

There is a general rule for the sum of those constants in every sector being
$3\pi/2$ (due to antiferromagnetic ground state), but excitations containing 
more than one magnon, e.g. $(2,+)$-strings, must be taken with their
appropriate multiplicity.

\section{The speed of sound in different sectors}
In this section we shall present some calculations for the speed of sound in 
sectors with one finite Fermi radius (the other one stays infinite) where this
radius is either small or large, so that an approximation can be made. On the 
other hand, that gives just the values of the speed near to the border lines of
the above mentioned sector which might be of interest for understanding the
physics of the model. We shall limit our calculations to the case $\bar{c}>0$,
$\tilde{c}<0$. Thus, the sectors in question are I and III. Sectors V and VI 
could be treated in an analogous way.

We start with sector I. The equations of the thermodynamic Bethe ansatz can be 
taken from \cite{doerfel3} (equation (4.5)).
\begin{eqnarray}\label{de1}
\epsilon_{1+}(\lambda) = - 2\pi\bar{c} s_1(\lambda)- 2\pi\tilde{c} d_1(\lambda)
+ d_1*\epsilon_{1+}^+(\lambda),
\end{eqnarray}
\begin{eqnarray}\label{de2}
\epsilon_{1-}(\lambda)= 2\pi\tilde{c}s_2(\lambda)-s_2*\epsilon_{1+}^+(\lambda).
\end{eqnarray}
Here $\varepsilon_i(\lambda)$ are the dressed energies. Their lower index 
specifies the excitation, while the upper stands for the positive and negative 
parts respectively. The convolution is defined by
\begin{equation}
a*b(\lambda) = \int_{-\infty}^{\infty}d\mu a(\lambda-\mu)b(\mu).
\end{equation}
(For $s_1(\lambda)$ and $s_2(\lambda)$ see the appendix.)

Restricting $\gamma$ by $0<\gamma<2\pi/5$ we consider the case
\begin{eqnarray}\label{case1}
0<\beta=\bar{c}-\frac{|\tilde{c}|}{2 \cos \tilde{\gamma}} \ll 1
\end{eqnarray}
with the usual definition $\tilde{\gamma}=\pi\gamma/2(\pi-\gamma)$.

From equation (3.13) from \cite{doerfel1} we know that $\beta=0$ describes the
border line between sectors IV and I.

To employ the method of power expansions in the (small) Fermi radius $b$ of the
$(1,+)$-strings, it is useful to reformulate equations (\ref{de1}) and 
(\ref{de2}) in the way
\begin{eqnarray}\label{de1re}
\varepsilon_{1+}(\lambda) = - \bar{c} g(\lambda,\pi-2\gamma) - \tilde{c} 
g(\lambda,\pi-3\gamma) - \frac{1}{2\pi} 
g(\lambda,\pi-3\gamma)*\varepsilon_{1+}^-(\lambda),
\end{eqnarray}
\begin{eqnarray}\label{de2re}
\varepsilon_{1-}(\lambda) = 2\pi \bar{c} s_1(\lambda) + \tilde{c} 
g(\lambda,\gamma) + \frac{1}{2\pi} 
g(\lambda,\gamma)*\varepsilon_{1+}^-(\lambda).
\end{eqnarray}
Equation (\ref{de1re}) determines the Fermi radius $b$, in lowest order the 
convolution term be neglected and $b$ is given by $\varepsilon_{1+}(b)=0$.
The derivative $d\varepsilon/dp$ is calculated the way (see equation (5.1) in
\cite{doerfel3})
\begin{eqnarray}\label{disp5}
\frac{d\varepsilon}{dp} = \frac{d\varepsilon}{d\lambda} \frac{d\lambda}{dp}
= - \left. \frac{d\varepsilon}{d\lambda} \frac{2}{\varepsilon(\lambda)} 
\right|_{\bar{c}=\tilde{c}=1}.
\end{eqnarray}
The speed of sound is then given if $\lambda$ is put equal to $b$.

After rather long technical calculations we have obtained
\begin{eqnarray}\label{sos5+}
v_{1+} = \frac{4\pi}{\pi-\gamma} \frac{1}{\sin \tilde{\gamma}} 
\sqrt{\frac{2\cos\tilde{\gamma}-1}{2\cos\tilde{\gamma}+1}}(\bar{c}\beta)^{1/2}.
\end{eqnarray}
Equation (\ref{de2re}) gives the same way
\begin{eqnarray}\label{sos5-}
v_{1-} = \frac{2\pi}{\pi-\gamma} (2\cos\tilde{\gamma}-1) \bar{c} +
\frac{8}{\pi-\gamma} \sin \tilde{\gamma} 
\sqrt{\frac{2\cos\tilde{\gamma}-1}{2\cos\tilde{\gamma}+1}}(\bar{c}\beta)^{1/2}.
\end{eqnarray}
Here the first term is just the limiting value on the border of sector IV.

Next we consider $\pi/3<\gamma<2\pi/5$ and look for the border of sector I with
sector II (see \cite{doerfel3}, equation (4.10)). It is given by
\begin{eqnarray}\label{border}
0<\alpha=|\tilde{c}| \tan\left(\frac{\pi^2}{2\gamma}\right)-\bar{c}\ll1.
\end{eqnarray}
Then equation (\ref{de1}) is of Wiener-Hopf type and can be solved explicitely
for large $b$ giving
\begin{eqnarray}\label{sos6+}
v_{1+} = \frac{2\pi}{\gamma} \frac{3\gamma-\pi}{\pi-2\gamma} 
\frac{\alpha}{1+ \tan\left(\frac{\pi^2}{2\gamma}\right)}.
\end{eqnarray}
From equation (\ref{de2}) one has to extract the asymptotics of 
$\varepsilon_{1-}(\lambda)$ for large $\lambda$ (Fermi radius is infinite).
The first term on the RHS gives the limiting value of $v_{1-}$ on the border of
sector II (it is non-vanishing). Remarkably for $\alpha>0$ the asymptotics is
always given by the second term with the convolution. So the result is of the 
same order in $\alpha$ as $v_{1+}$:
\begin{eqnarray}\label{sos6-}
v_{1-} = \frac{2\pi}{\gamma} \frac{3\gamma-\pi}{\gamma} 
\frac{\alpha}{1+ \tan\left(\frac{\pi^2}{2\gamma}\right)}.
\end{eqnarray}
Therefore $v_{1-}$ shows a discontinuity on the border I/II.

After sector I we consider sector III. Then one has $2\pi/5<\gamma<\pi/2$. We
start again with the border to sector IV.
\begin{eqnarray}\label{case?}
0<\beta'=\bar{c}-2 \cos \tilde{\gamma}\tilde{c} \ll 1.
\end{eqnarray}
The Fermi radius is determined again from equation (\ref{de1re}), but now 
instead of power expansion we have to solve a Wiener-Hopf problem due to the 
fact that $b$ is large. The calculation is straightforward after realizing that
for the positive frequency part of the inhomogenous function one has to take 
into account the dominant two poles (instead of one). As already mentioned in
\cite{doerfel3} the fact that filling of the $(1,+)$-strings starts at infinity
causes the appearance of two velocities for them taken at $\lambda=b$ and 
$\lambda=\infty$. While the calculations for both are different the results 
coincide in first order.

The same happens for $v_{1-}$ calculated from the asymptotics of equation 
(\ref{de2re}). We remind the reader that $v_{1-}$ vanishes at the border. 
Finally we have obtained
\begin{eqnarray}\label{sos_all}
v_{1+}^{(1)} = v_{1+}^{(2)} = v_{1-} = \frac{2\pi}{\pi-\gamma} 
\frac{\beta'}{1+2 \cos \tilde{\gamma}}.
\end{eqnarray} 
We have no explanation for that striking fact, that all velocities are equal in
first order. Nevertheless we are sure that the effect does not survive in 
higher order.

To close that part we have to consider the border line with sector II (see 
\cite{doerfel3}, equation (4.8)).
\begin{eqnarray}\label{alpha_prime}
0<\alpha'=|\tilde{c}| 2\gamma d_1(0) - \bar{c} \ll 1.
\end{eqnarray}
The Fermi radius $b$ is determined by equation (\ref{de1}) using power 
expansion. The calculation is straightforward, but rather tedious because of 
the presence of the function $d_1(\lambda)$. But the order of magnitude can be 
estimated easily. The two velocities $v_{1-}^{(1)}$ and $v_{1+}^{(2)}$
(taken at 
$\lambda=\infty$) obtain corrections to their (constant) limiting values from
sector II of the order $\sqrt{\alpha'}$. The third velocity $v_{1+}^{(1)}$
(taken at $\lambda=b$) is of order $\sqrt{\alpha'}$.

We did not consider the border of sectors I and 0 for $0<\gamma\leq\pi/3$ 
because both sectors are separated form each other by a singular line with a 
highly degenerate ground state which is different from the other lines 
considered above.

\section{Conclusions}
First we make some general comments on the character of the behaviour of the 
velocities considered. We parametrize the coupling constants by two parameters,
one $c$, setting the energy scale (below put to one), the other one an angle 
$\varphi$, which describes their ratio.
\begin{eqnarray}\label{cphi}
\bar{c}=c\sin\varphi,\nonumber\\
\tilde{c}=c\cos\varphi.
\end{eqnarray}
In general from our results in section 4 one can see that the behaviour near 
the sector borders is either linear or a square root in cosine/sine-functions
of $\varphi$. The fact which one is chosen is determined by the radius $b$. 
Large $b$ produces Wiener-Hopf equations and linear behaviour, small $b$ causes
power expansions and a square root behaviour. Considering sector I only, one 
could superficially conclude that this depends only on the sector on the other 
side of the border. Our analysis in sector III shows that the situation 
reverses with respect to sector I.

\begin{figure}
\psfrag{v11111}{$v_{1+}$}
\psfrag{v22222}{$v_{2+}$}
\psfrag{v33333}{$v_{1-}$}
\includegraphics[width=\columnwidth]{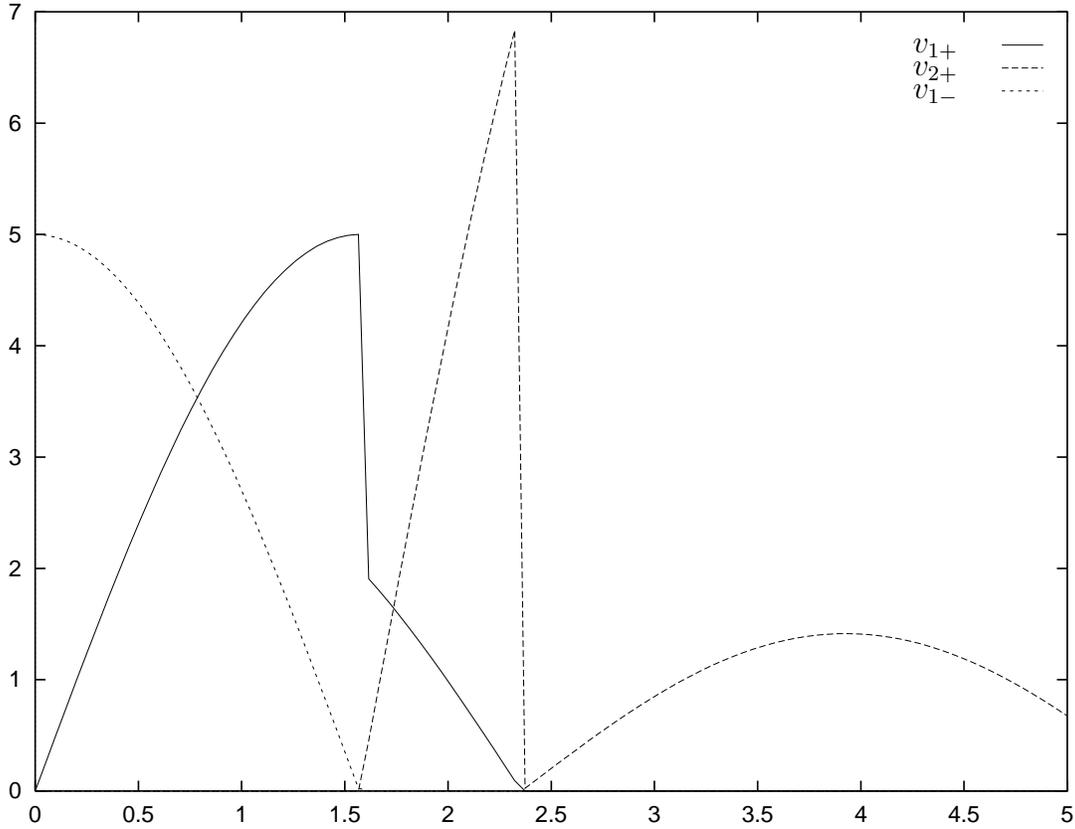}
\caption{\label{g2p5}The speeds of sound as functions of the angle
$\varphi$ for $\gamma=2\pi/5$ and those sectors where all speeds of
sound are known.}
\end{figure}

In just one case we have found a discontinuity of one of the velocities when a
sector border is passed (see equation (\ref{sos6-})). It is the case for $v_{1-}$ 
on the border I/II. Nothing of this kind happens for sector III, because for 
small $\alpha'$ we have power expansion, non-capable of producing such an 
effect. On the other hand (for small $\beta'$) $v_{1-}$ vanishes on the other
side of the border in sector IV. In general, one would expect a discontinuity
only when a singular line is crossed which means that an excitation with
infiniteFermi zone disappears at once (see below). It is rather instructive to 
compile all analytically calculated velocities in one picture at the special 
point $\gamma=2\pi/5$ which we have done in figure \ref{g2p5}.

For convenience, we have given the analytical shape in the various sectors in 
table \ref{g2p5t}.
\begin{table}
\begin{center}
\begin{tabular}{|c||c||c||c||c|}\hline
sectors & 0 & II & IV & V\\ \hline
borders & $0$, $\pi/2$ & $\pi/2$,$3\pi/4$ & $3\pi/4$, $2\pi-\arctan 4$ &
$2\pi-\arctan 4$, $2\pi$ \\ \hline
$v_{1+}$ & $5\sin\varphi$ & $5/2\sin\varphi+5/2\cos\varphi$ & 0 & 0 \\ \hline
$v_{2+}$ & $5\cos\varphi$ & 0 & 0 & \\ \hline
$v_{1-}$ & 0 &  $-10\cos\varphi$ & $-5/3\sin\varphi-5/3\cos\varphi$ & \\ \hline
\end{tabular}
\end{center}
\caption{\label{g2p5t}The speeds of sound and the sector borders as functions 
of the angle $\varphi$ for $\gamma=2\pi/5$.}
\end{table}
Looking at figure \ref{g2p5} one clearly sees two effects. First the two 
singularity lines at $\varphi=\pi/2$ (the vanishing of the $(2,+)$-strings)
and at $\varphi=3\pi/4$ (the vanishing of the $(1,+)$-strings) cause 
discontinuities for the speed of the remaining excitation. While the 
first line is universal for arbitrary $\gamma$, the second one is peculiar to 
$\gamma=2\pi/5$. For other $\gamma$ that line is replaced by sectors I or III.

Second one sees the two conformal points where two velocities intersect, one at
$\varphi=\pi/4$ and the other at $\varphi=\pi-\arctan 5$ \cite{doerfel1}.

Our calculations in the previous section allow now to look closer at what 
happens for $\pi/3<\gamma<2\pi/5$ when the above mentioned singularity line is 
replaced by sector I. As far as the limits of sectors II and IV are concerned 
the only change is that $v_{1-}$ no longer vanishes at the border of sector IV.
Now we make use of equations (\ref{sos5-}), (\ref{border}),(\ref{sos6+}) and 
(\ref{sos6-}). They tell us first that both velocities must reach their
maximum (at different $\varphi$) within sector I and second that there
is (at least one) point of
intersection in the same sector. In that point the finite size corrections must
exhibit a rather simple structure like in \cite{woy}.

We conjecture that there is no such effect in sector III but a strict prove 
requires further calculations. In addition, our calculations show another 
interesting effect, at $\beta'=0$ all velocities vanish.

Finally we remark that we expect the theory at $\gamma=2\pi/5$ and both 
velocities equal to be highly symmetric and of particular interest. Up to now 
it is only known that its central charge is $2$. We were not able to calculate 
the operator dimensions analytically, because contrary to what happened in case
of paper \cite{doerfel2} it requires the knowledge of the dressed charge 
\cite{woy} and hence the explicit factorization of the kernel.

\section*{Acknowledgment}
The authors would like to thank H. Frahm for helpful discussions.

\section*{Appendix}
\begin{eqnarray*}
s_1(\lambda)=\frac{1}{2\gamma\cosh(\pi\lambda/\gamma})\\
s_2(\lambda)=\frac{1}{2(\pi-2\gamma)\cosh(\pi\lambda/(\pi-2\gamma))}\\
g(\lambda,\alpha)=\int_{-\infty}^{\infty} e^{i\omega\lambda}
\frac{\cosh\omega\alpha/2}{\cosh\omega(\pi-\gamma)/2} d\omega
\end{eqnarray*}

\newpage
\section*{Figure and table captions}
{\bf Figure 1.} Dispersion relations for the holes in the 
$(1,+)$-string distributions in sector II for $\bar{c}=-5\tilde{c}=1$ and 
different $\gamma$.\\[2cm]
{\bf Figure 2.} The speeds of sound as functions of the angle $\varphi$ for 
$\gamma=2\pi/5$.\\[2cm]
{\bf Table 1.} All sectors appearing in the phase diagram. 
Upper indices indicate infinite and finite Fermi zones. In the latter case, 
the second index distinguishes, wether the filling starts at $\lambda=0$ or 
$\lambda=\infty$.\\[2cm]
{\bf Table 2.} The speeds of sound and the sector borders as functions of the 
angle $\varphi$ for $\gamma=2\pi/5$.


\section*{References}
\begin{thebibliography}{99}
\bibitem{doerfel3}D\"orfel B - D and Mei\ss ner St 1998
                 {\it J. Phys. A: Math. Gen.} {\bf 31} 61
\bibitem{devega}de Vega H J and Woynarovich F 1992
                 {\it J. Phys. A: Math. Gen.} {\bf 25} 4499
\bibitem{meissner}Mei\ss ner St and D\"orfel B - D 1996
                 {\it J. Phys. A: Math. Gen.} {\bf 29} 1949
\bibitem{doerfel1}D\"orfel B - D and Mei\ss ner St 1996
                 {\it J. Phys. A: Math. Gen.} {\bf 29} 6471
\bibitem{doerfel2}D\"orfel B - D and Mei\ss ner St 1997
                 {\it J. Phys. A: Math. Gen.} {\bf 30} 1831
\bibitem{woy}Woynarovich F 1989
                 {\it J. Phys. A: Math. Gen.} {\bf 22} 4243
\end{thebibliography}
\end{document}